\documentclass[9pt,twocolumn,twoside]{osajnl}

\pdfoutput=1

\usepackage{graphicx}
\usepackage{wasysym}
\usepackage{epstopdf}

\usepackage{color}

\journal{ao} 

\setboolean{shortarticle}{false} 

\ifthenelse{\boolean{shortarticle}}{\colorlet{color2}{color2b}}{\colorlet{color2}{color2}} 

\title{A medium-finesse optical cavity for the stabilization of Rydberg lasers}
\author[1,*]{Julius de Hond}
\author[1]{Nataly Cisternas}
\author[1,$\circ$]{Graham Lochead}
\author[1]{N.J. van Druten}

\affil[1]{Van der Waals -- Zeeman Institute, Institute of Physics, University of Amsterdam, Science Park 904, 1098XH Amsterdam, The Netherlands}
\affil[$\circ$]{Present address: Physikalisches Institut, Universit\"at Heidelberg, Im Neuenheimer Feld 226, 69120 Heidelberg, Germany}

\affil[*]{Corresponding author: j.j.m.dehond@uva.nl}

\dates{Compiled \today}

\ociscodes{(020.5780) Rydberg states; (120.2230) Fabry-Perot; (140.3425) Laser stabilization; (300.3700) Linewidth; (300.6310) Spectroscopy, heterodyne}

\doi{}

\begin{abstract}
We describe the design, construction, and characterization of a medium-finesse Fabry--P\'erot cavity for simultaneous frequency stabilization of two lasers operating at 960 and 780~nm wavelengths, respectively. The lasers are applied in experiments with ultracold rubidium Rydberg atoms, for which a combined laser linewidth similar to the natural Rydberg linewidth ($\approx 10~\mathrm{kHz}$) is desired. The cavity, with a finesse of $\approx 1500$, is used to reduce the linewidth of the lasers to below this level. By using a spacer made of ultra low expansion (ULE{\textsuperscript\textregistered}) glass with active temperature stabilization, the residual frequency drift is limited to $\lesssim 1~\mathrm{MHz/day}$. The design optimizes for ease of construction, robustness, and affordability.
\end{abstract}

\setboolean{displaycopyright}{true}

\begin{document}
\maketitle
\thispagestyle{fancy}

\ifthenelse{\boolean{shortarticle}}{\ifthenelse{\boolean{singlecolumn}}{\abscontentformatted}{\abscontent}}{}

\section{Introduction}
Ever since the original experiments by Fabry and P\'erot, optical cavities \cite{Siegman86,Kogelnik66} have been useful tools for accurately measuring distances and creating frequency-specific light sources \cite{Perot99}. Nowadays their uses are versatile, and they can be found in numerous places where stable and/or narrow wavelength lasers are required. This includes but is not limited to their use in the LIGO facilities for the detection of gravitational waves \cite{Mueller16}; their relevance as a source for stabilizing lasers with sub-Hz linewidths \cite{Alnis08,Ludlow07,Zhang13,Wu16}; and their use as frequency-selective elements for wavelength multiplexing in optical communication networks \cite{Brackett90}. Cavities are popular in the field of atomic physics, too, as a source for stabilization of lasers, and as a platform for exploring strong light--matter couplings, for example, in cavity QED \cite{Fox06, Haroche12}.

In the 1960's cavities were predominantly used for stabilizing gas and dye lasers \cite{White65,Barger73,Drever83}, but since the advent of semiconductor lasers they have proven to be a reliable source for stabilizing them, too. Because of a diode's sensitivity to optical feedback, it is for instance possible to stabilize it by feeding the reflected signal of the cavity directly back into the diode \cite{Dahmani87}. Nowadays, however, with the advent of cheap and high-bandwidth feedback electronics, it is more common to generate an error signal which is fed back into the diode electronically \cite{Yamamoto85,Ohtsu85}. Although there exist several schemes for generating an electronic error signal, the most common technique is that of Pound--Drever--Hall (PDH) stabilization, where the phase of the laser is modulated with an RF signal \cite{Bjorklund83,Black01}.

Cavities prove to be an excellent source for stabilization signals in a wide range of atomic physics experiments, for instance that of Rydberg physics. Rydberg atoms \cite{Gallagher94} have lifetimes in the $100~\mu$s range (depending on principle quantum number), and therefore their optical transition linewidths are typically in the kHz regime. Several proposals, for instance Rydberg dressing \cite{Balewski13-2,Johnson10}, require narrowed lasers with linewidths on the order of $10~\mathrm{kHz}$. Cavities are convenient in this context, because through the choice of mirrors the width of the reference feature can be engineered \cite{Kogelnik66,Siegman86}---as opposed to Doppler-free saturation spectroscopy in vapor cells, typically used for laser cooling applications, where the linewidth is limited by the natural linewidth of the relevant atomic species (for rubidium this is several MHz \cite{Steck08}). In a two-photon excitation scheme vapor cells can in principle also be used for locking the upper transition's laser, e.g.\ using a method based on electromagnetically induced transparency, for instance \cite{Abel09}. This requires a large amount of laser power, and it also means the laser frequency cannot be scanned without additional steps. Finally, a cavity is advantageous compared to atomic transitions because it does not saturate.

A disadvantage of a locking scheme based on a cavity is that the resonance frequencies are arbitrary and are not in any way related to atomic transitions. This is a problem that can be mitigated by a locking scheme that relies on a stable fixed-length cavity combined with movable sidebands \cite{Thorpe08}. 

Here we describe the design, construction, and characterization of a medium-finesse cavity setup and its use for linewidth narrowing and stabilizing two diode lasers that are used for (two-photon) Rydberg excitation of ultracold rubidium gases.

Our linewidth requirements are at the kHz level where a medium-finesse cavity suffices. Our design was optimized for affordability, ease of construction, and robustness. It is intended for two-photon excitation of high-lying Rydberg states ($n \geq 20$) from the ground state of $^{87}$Rb using lasers operating at $780~\mathrm{nm}$ (near the $5s_{1/2}$--$5p_{3/2}$ transition) and at $480~\mathrm{nm}$ (near the transition from the $5p_{3/2}$ level to $ns$ or $nd$ Rydberg levels). The cavity that is described here was designed to operate at wavelengths of $780$ and $960~\mathrm{nm}$ (the $480~\mathrm{nm}$ light that was used for Rydberg excitation was derived from the latter using second harmonic generation). This is similar to the setup described in Ref.~\cite{Naber16} by Naber \emph{et al.}~where they employed a commercial cavity from Stable Laser Systems \cite{SLSdata} for a similar purpose. Our experiment also bears resemblance to that of L\"ow \emph{et al.}~described in Ref.~\cite{Low12}. Here a {Zerodur\textsuperscript\textregistered} spacer block with two separate bores---one for each wavelength---was used. The length of each cavity was independently controllable with a special piezo stack that is insensitive to thermal fluctuations \cite{Low12}.) Their cavity effectively consists of two cavities, and is more complicated to assemble than ours. The design presented here is inspired by that of Stellmer as described in Ref.~\cite{Stellmer13} and that of Boddy from Ref.~\cite{Boddy14}.

\section{Design considerations}\label{sec:design-considerations}
The design principle of most cavities is to attach two highly-reflective mirrors to a spacer block made of material that has a stable length. Often this is a material with a low thermal expansion coefficient, but the length can also be actively stabilized. This spacer, then, is placed in a stable mount which is put in a vacuum chamber. Our design incorporates some improvements over Stellmer's design, some of which were suggested in Ref.~\cite{Stellmer13}, which will be elaborated below.

For the design of our cavity several considerations were taken into account. First of all, to prevent air currents and short-term thermal fluctuations from changing the alignment or the spacer length, the cavity is mounted in a vacuum system. This is pumped by an ion pump to avoid mechanical vibrations. Because we wanted to implement a dual laser-locking scheme with the same cavity, a piezo-based design was less suitable. That is because it is not possible to independently control the resonance positions of two different lasers this way. Additionally, a piezo can cause mechanical instabilities and requires feeding through wires through the vacuum system, which makes a sideband-locking scheme all the more attractive \cite{Thorpe08}.

Our cavity is non-hemispherical, and the mirrors are in a plano-concave configuration; the reason for which is threefold: i) The alignment of the mirrors' optical axes in a plano-concave cavity is not crucial, making the assembly easier, ii) a confocal configuration puts stringent constraints on the spacer length, and iii) in a realistic confocal system the higher order modes will not perfectly overlap with the principal mode due to small misalignments, leading to artificial broadening \cite{Siegman86}. 

The reflectivity of the mirrors we need is set by requirements on the finesse $\mathcal{F}$ of the cavity: this is the ratio of the free spectral range ($f_\mathrm{FSR}=c/2L$) and the full width at half maximum (FWHM) $\delta\!f$ of the principal cavity resonances.

Fixing the finesse and the cavity length $L$ thus fixes the width of the resonances, which in turn determines the possible linewidths that can be achieved by locking to it. Using a typical PDH setup it is possible to reduce the laser linewidth to a fraction of the cavity linewidth \cite{Black01,Bjorklund83}. Reduction by factors of $10^{-4}$ is typical, and it is even possible to achieve a factor of $10^{-6}$ \cite{Zhang14}. For our experiments, we aim for a factor better than $10^{-2}$. Since the lasers have to be narrowed to $\sim 10~\mathrm{kHz}$ this requires a cavity linewidth of $\sim 1~\mathrm{MHz}$. Furthermore, the cavity should not be too long, since this will make it more sensitive to vibrations \cite{Alnis08,Ludlow07}. It must also not be too short, for then the free spectral range becomes so large it cannot be spanned by the movable sidebands. We choose a length $L = 100~\mathrm{mm}$, which puts the required finesse at $\left(c/2L\right)/\delta\!f \approx 1500$.

Last but not least there is the price of the system. Commercial cavities of a similar design made by Stable Laser Systems are used elsewhere in the field (see e.g.\ Ref.~\cite{Naber16}). By using a design that combines two low-expansion glasses ({ZERODUR\textsuperscript\textregistered} and {ULE\textsuperscript\textregistered} \cite{SLSdata}), the typical finesses and long-term drifts of their cavities are better than the design we are describing presently. There is a price tag attached, though; our complete setup costs less than €7,000. This includes the vacuum system with pump, and the production of the spacer and mirrors; the fabrication of the latter is the most expensive, since the mirrors have to be highly reflective for both 780 and 960~nm. A fully assembled cavity from Stable Laser Systems (SLS) costs several times that \cite{SLSdata}.

\section{The cavity design}\label{sec:cavity-design}
The cavity itself consists of a spacer of ultra-low expansion (ULE\textsuperscript\textregistered) glass (measuring $(30\pm 0.1) \times (30 \pm 0.1) \times (100 \pm 0.2)~\mathrm{mm^3}$), which was custom made by Hellma Optik. The reason for choosing {ULE\textsuperscript\textregistered} was that it has an extremely low coefficient of thermal expansion ($0 \pm 30 \times 10^{-9}/^\circ\mathrm{C}$), which has a zero crossing close to room temperature \cite{ULEData}. Alternatives such as {ZERODUR\textsuperscript\textregistered} have slightly higher expansion coefficients but may be better at other temperatures \cite{ZerodurData}. The spacer has a main bore centered along the long axis ($\diameter 10~\mathrm{mm}$) and an additional bore ($\diameter 3~\mathrm{mm}$) in one of the sides which is intended for venting.

The cavity mirrors were custom made by Layertec GmbH. The radii of curvature were chosen as $R_1=500~\mathrm{mm}$ and $R_2=\infty$, making this a plano-concave cavity well within the stability region \cite{Kogelnik66,Siegman86}, given by the criterion
\begin{equation}
0 \leq g_1 g_2 \leq 1,
\end{equation}
with
\begin{equation}
g_i=1-\frac{L}{R_i}, \notag
\end{equation}
where $L$ is the length of the cavity.

Additionally, for these values the resonance frequencies of the first few higher-order transverse modes are different from those of the principal modes.

The transverse mode spacing is a fraction $\arccos\left(\sqrt{g_1 g_2}\right)/\pi \approx 0.15$ of the free spectral range \cite{Siegman86}, so that the first few transverse modes will be clearly distinguishable from the principal ones.

The mirrors are specified to have a reflectivity of $99.8 \pm 0.1 \%$ at wavelengths of $780~\mathrm{nm}$ and $958\mbox{--}970~\mathrm{nm}$, leading to a finesse \cite{Siegman86}
\begin{equation}\label{eq:Finesse-ideally}
	\mathcal{F} = \frac{\pi\sqrt{R}}{1-R} \approx 2000.
\end{equation}

The mirrors were glued onto the spacer using {EPO-TEK\textsuperscript\textregistered} 353ND epoxy \cite{NASAOutgassing}. This was chosen for its low outgassing rates such that it is suitable for use in a vacuum system.

The assembly of the spacer and the mirrors is placed in a hemicylindrical block of stainless steel with a gutter to accommodate the spacer. This gutter was designed to have a $2^\circ$ angle with respect to the long axis of the block to prevent any problems with back reflections from the windows of the vacuum assembly in which it is placed. The spacer is supported by four short {Viton\textsuperscript\textregistered} rods (with a length and diameter of approximately $20$ and $5~\mathrm{mm}$, respectively).

\begin{figure}
	\centering
	\centering
	\includegraphics[width=0.5\textwidth]{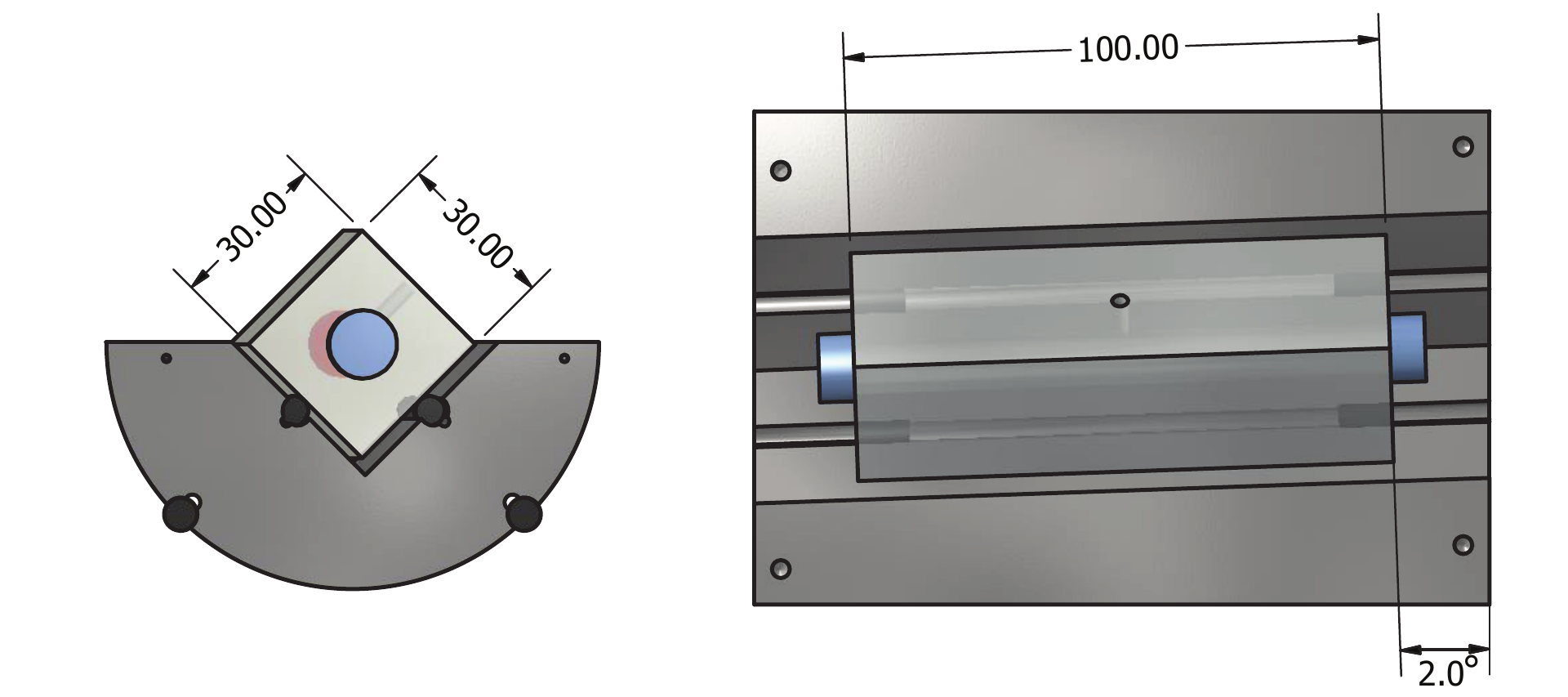}
	\includegraphics[height=3cm]{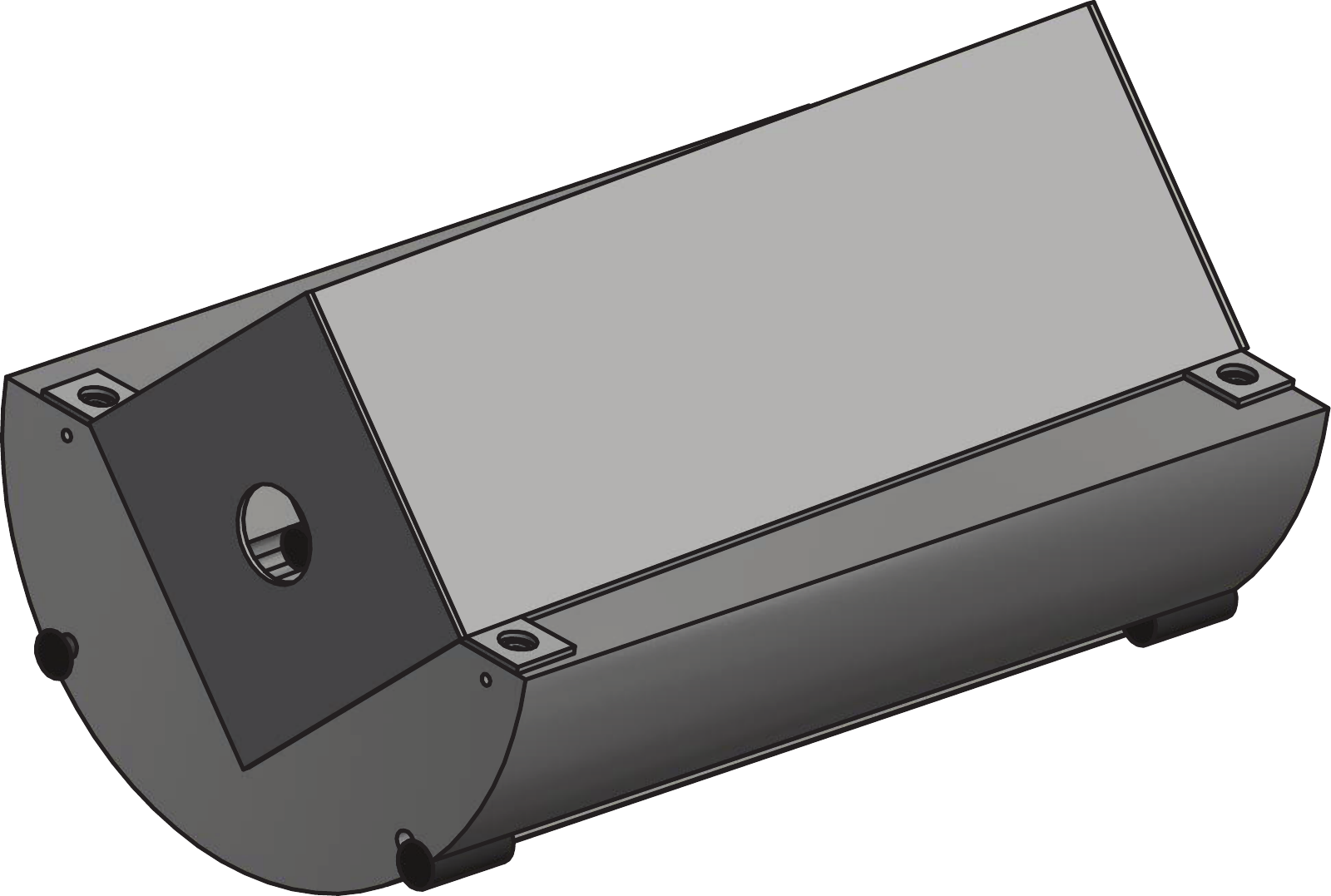}
	\caption{Drawings of the cavity design. Top: Exposed view from front (left) and top (right) with relevant dimensions in mm or degrees. Bottom: View with the cover included.}
	\label{fig:cavity-renderings}
\end{figure}

A cover was designed to protect the exposed part of the spacer from black-body radiation and hazards such as debris coming from the ion pump (to which there is a direct line-of-sight otherwise). This cover has two holes on either side for optical access to the cavity. Figure~\ref{fig:cavity-renderings} includes some renderings of the full design. In Stellmer's original cavity, the spacer was enclosed, and clamped in tight by a cylinder of stainless steel and thereby protected \cite{Stellmer13}. Despite this being more rigid, it does cause stresses on the spacer that are not desirable.

\section{The vacuum system}\label{sec:vacuum-system}
\begin{figure}
	\centering
	\includegraphics[width=0.45\textwidth]{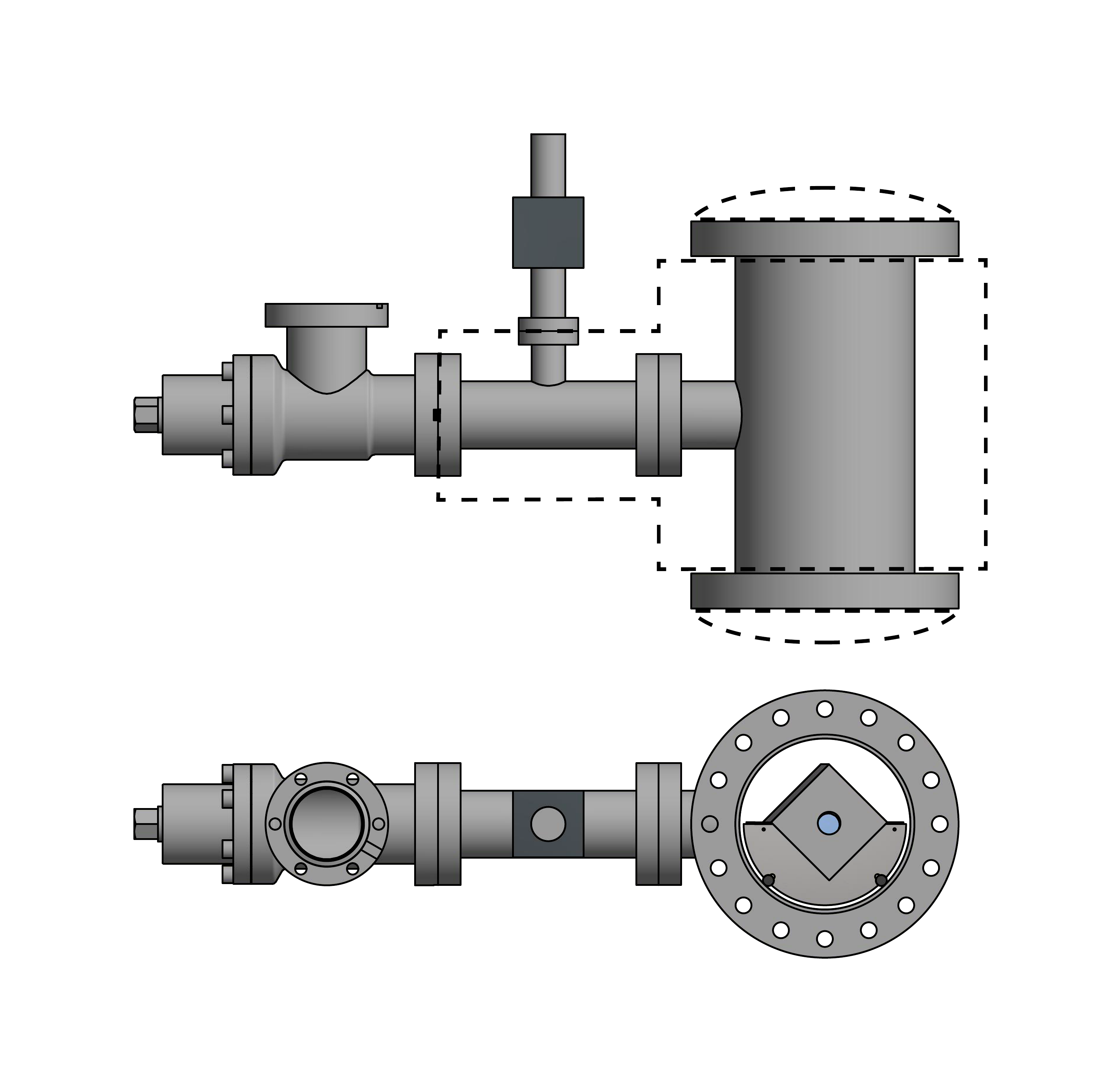}
	\caption{Drawings of the vacuum system with the cavity mounted inside. Top: top view of the system. The dashed lines indicate the areas that are wrapped in Superwool{\textsuperscript\textregistered}. Bottom: side view of the system. In both drawings the ion pump is in the center (represented by the black square) and the valve is on the left.}
	\label{fig:vacuum-system}
\end{figure}

The stainless steel block was placed in a vacuum system to reduce temperature and air fluctuations. It was also supported by short {Viton\textsuperscript\textregistered} rods (of similar dimensions as those mentioned above) so that it does not contact the vacuum chamber directly. A diagram of the assembly can be seen in Fig.~\ref{fig:vacuum-system}.

The system consists of standard vacuum parts: two reducing tees (DN100 to DN40 and DN40 to DN16); the mount is placed in the larger tee, and the smaller one is used to attach an ion pump and a valve. There are viewports on either side of the larger tee, allowing us to monitor both the reflection and the transmission signals. The flatness of standard viewports can be a problem, however, viewports that are suitable for laser applications are expensive. Therefore we chose to equip the chamber with two different viewports: one made from standard {Kodial\textsuperscript\textregistered} glass, and one made from UV grade fused silica. They were used on the transmission and reflection sides of the cavity, respectively, because it is the reflection signal we are interested in for locking.

The system has a small, dedicated ion pump (Agilent, $2~\mathrm{l/s}$) for maintaining vacuum conditions. There is also a valve which shuts off a flange that can be used for connecting a turbo pump; this is only used for initially evacuating the system, and not part of the day-to-day operation. After the system was assembled, it was evacuated using a turbo pump, and baked at a modest temperature of $50~^\circ\mathrm{C}$ for one week.

Subsequently the valve was closed and the turbo pump was disconnected, leading to a final pressure of about $3\times10^{-7}~\mathrm{mbar}$ (as derived from the ion pump current).

\section{Temperature stabilization}\label{sec:temp-stabilization}
Despite our best efforts to insulate the cavity from the environment by placing it in a vacuum system, temperature fluctuations are bound to couple into it. Small changes on the order of $100~\mathrm{mK}$ will affect the length of the cavity slightly and thereby change the resonance positions. We will now estimate the size of this effect. Any principal cavity mode satisfies $f_n = nc/2L$, with $n$ the mode index. If the length $L$ of the spacer changes by an amount $\Delta L$, this means that the resonance frequency will shift by
\begin{equation}\label{eq:freq-shift}
	\Delta f_n = \frac{nc}{2L} - \frac{nc}{2\left(L+\Delta L\right)}.
\end{equation}
The change in length is determined by the coefficient of thermal expansion $\alpha_T$ through $\Delta L = \alpha_T L \Delta T$. Using this, and setting $f_n = c/\lambda$ (where $\lambda$ is the wavelength of the light) we get
\begin{equation}\label{eq:freq-shift-expansion-coeff}
	\Delta f_n = \frac{c}{\lambda} \frac{\alpha_T\Delta T}{1+\alpha\Delta T}\approx \frac{c\alpha\Delta T}{\lambda},
\end{equation}
where in the last step we used $\alpha\Delta T \ll 1$.

For $\lambda= 780~\mathrm{nm}$ and $\alpha = 30\cdot 10^{-9}~\mathrm{K}^{-1}$ (the upper limit to the thermal expansion coefficient of ULE{\textsuperscript\textregistered} \cite{ULEData}) this means that a temperature change of $100~\mathrm{mK}$ will result in a length change of $0.30~\mathrm{nm}$, corresponding to a frequency shift of $1.15~\mathrm{MHz}$. Without any additional stabilization, such fluctuations are likely to find their way into the system, because the laboratory temperature is only regulated at the level of $1~\mathrm{K}$.

In order to stabilize the temperature and thereby stabilize the cavity itself, a resistive heating pad\footnote{Thermo Fl\"achenheizungs GmbH, measuring $230~\mathrm{mm}\times 130~\mathrm{mm}$ and specified to provide $20~\mathrm{W}$ at $24~\mathrm{V}_\mathrm{AC/DC}$.} 
was stuck to the DN100 part of the large reducing tee. Three temperature sensors (LM35, Texas Instruments, Inc.) and two thermistors ($10~\mathrm{k}\Omega$, NTC, Epcos AG) were attached as well in order to monitor and regulate

the temperature, respectively.
For further insulation, the assembly was wrapped in a Superwool{\textsuperscript\textregistered} blanket \cite{Superwoolspec} which, in turn, was covered using aluminium foil.

To regulate the temperature we used a PTC5K-CH controller (Wavelength Electronics, Inc.) which is compatible with several temperature sensors including our $10~\mathrm{k}\Omega$ thermistor \cite{TempControlData}. It sends $100~\mu\mathrm{A}$ through the thermistor, and uses the output voltage as an error signal in a PI-circuit. For stable regulation, it was necessary to increase the time constant of the integrator by replacing the default $3.3~\mu\mathrm{F}$ capacitor by a $10~\mu\mathrm{F}$ capacitor. 

To reduce the coupling in of thermal fluctuations from the environment, we added a second temperature controller with heating foil that regulates the temperature of the mounts of the vacuum system enclosing the cavity. The foil\footnote{Thermo Fl\"achenheizungs GmbH, measuring $170~\mathrm{mm}\times 70~\mathrm{mm}$ and specified to provide $30~\mathrm{W}$ at $24~\mathrm{V}_\mathrm{AC/DC}$.} for this circuit was glued on the breadboard between the mounts, and the thermistor was placed on one of them.

The controller was optimized such that the temperature is stable and there is no crosstalk between the two controllers.

A change in temperature of the vacuum chamber will not result in an instantaneous equivalent temperature change of the steel mounting block and the cavity spacer. The heat flow between the components is resistively coupled via the {Viton\textsuperscript\textregistered} rods, and can be thought of as the thermal equivalent of an electrical RC-circuit. Here the {Viton\textsuperscript\textregistered} acts as thermal resistor, and the mount and spacer are both thermal capacitors. The parameters of interest are the heat conductivity and capacity of the different materials. First we will consider the heat flow between the vacuum housing and the steel mounting block.

The specific heat of steel is $0.483~\mathrm{J/gK}$ \cite{HBCP95}; with our mounting block weighing approximately $2~\mathrm{kg}$, this means that its heat capacity is $\sim 1~\mathrm{kJ/K}$. The precise thermal conductivity of {Viton\textsuperscript\textregistered} is unknown, it is a fluoroelastomer, which is a type of synthetic rubber. For our estimate we will use the conductivity of natural rubber, which is approximately $0.2~\mathrm{W/Km}$ \cite{HBCP95}. Under the assumption that the rods are cuboids this puts the thermal resistance at about $250~\mathrm{K/W}$. Since there are four rods connected in parallel, the total thermal resistance is a quarter of this number, so about $60~\mathrm{K/W}$. For the coupling of environmental fluctuations into the steel mount we thus find an RC-time of approximately $60~\mathrm{K/W} \cdot 1~\mathrm{kJ/K} \approx 17~\mathrm{hrs}$.

Similarly, the {ULE\textsuperscript\textregistered} spacer has a specific heat of $1.31~\mathrm{J/gK}$ \cite{ULEData} and a mass of $200~\mathrm{g}$, resulting in a heat capacity of $153~\mathrm{J/K}$. Assuming the same thermal resistance for the support rods this yields an RC-time of $\approx 3~\mathrm{hrs}$. This timescale is shorter than that of the steel mount, so heat transport into the mount will be the rate limiting step for heating of the whole assembly.

\section{RF Generation}\label{sec:rf-generation}
In order to control the frequency of the lasers and stabilize them in a spectroscopically relevant place, RF signals were needed for sideband generation. The sidebands were applied using fiber-based electro-optic modulators (EOMs) (PM785 and PM940 for the $780$ and $960~\mathrm{nm}$ light, respectively) from Jenoptik AG. For each laser, two RF signals were required: one to generate movable sidebands (in the $100$--$1500~\mathrm{MHz}$ range) to which the lasers could be locked, and one to generate sidebands at $\sim 25~\mathrm{MHz}$ for the Pound--Drever--Hall locking scheme \cite{Bjorklund83,Black01}.

The $25~\mathrm{MHz}$ signals were taken from the Toptica PDD110 modules of the respective lasers. The movable sidebands were generated using a controllable RF source. We used a SynthUSBII from WindFreak Technologies, LLC., which is an implementation of the ADF4351 chip from Analog Devices, Inc. It has a tuning range from $34~\mathrm{MHz}$--$4.4~\mathrm{GHz}$ and features an integrated microcontroller which makes it easy to incorporate in the existing laboratory software.

\section{Cavity analysis}\label{sec:cavity-analysis}
Using the $960~\mathrm{nm}$ light, spectra of the cavity were obtained. Using a scan over a full free spectral range we found this to equal $f_\mathrm{FSR} = 1495\pm 2~\mathrm{MHz}$. This frequency was calibrated by overlapping the movable sidebands of two adjacent modes. This free spectral range corresponds to $L=100.2\pm 0.1~\mathrm{mm}$, which is close to the specified length of the spacer.

Theoretically we expect $f_\mathrm{FSR} \approx 1500~\mathrm{MHz}$ for our case. The slight discrepancy between this and what we measure can (partially) be explained by the fact that one of the two mirrors is curved, effectively increasing the length of the cavity. Along the central axis the increase in length equals
\begin{equation}
	\Delta L = \frac{d^2}{4R_1}\frac{1}{\sqrt{1-\left( d/2R_1 \right)^2}} = 80.7~\mu\mathrm{m},
\end{equation}
(with $d = 12.7~\mathrm{mm}$ being the mirror's diameter and $R_1 = 500~\mathrm{mm}$ its radius of curvature). This means that now $f_\mathrm{FSR} = 1498~\mathrm{MHz}$. This is still not equal to what we measured, but it does put the length of the spacer within its specified margin.

\begin{figure}
	\centering
	\includegraphics[width=0.45\textwidth]{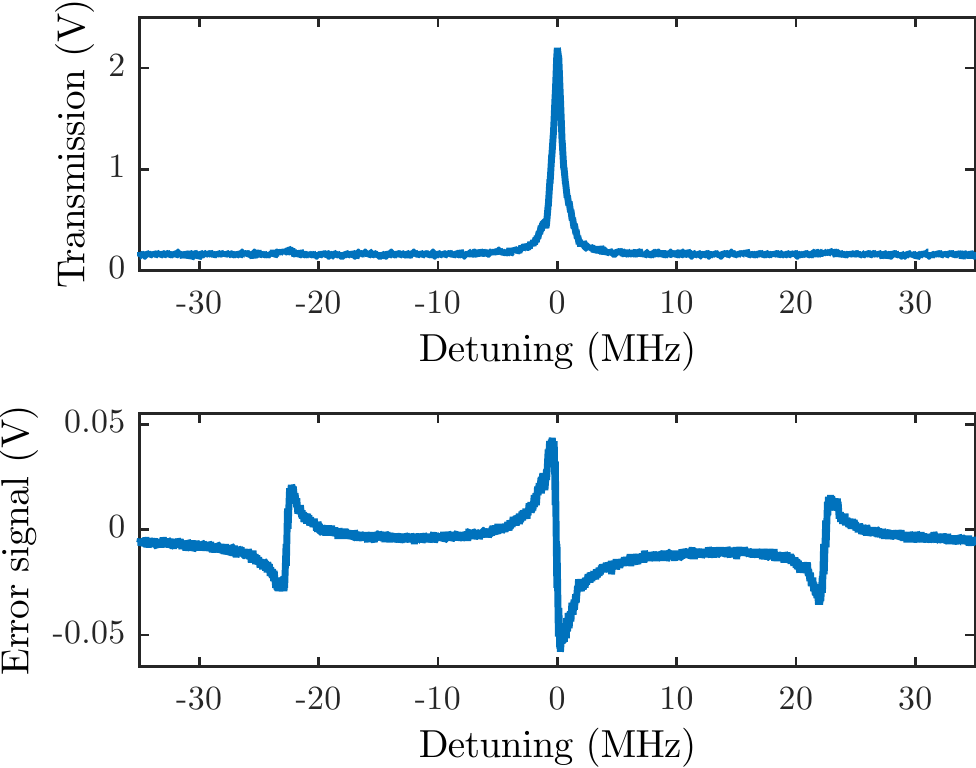}
	\caption{Top: Transmission spectrum of a principal mode of the $960~\mathrm{nm}$ beam with weak phase modulation sidebands at $\sim 25~\mathrm{MHz}$. Bottom: Error signal derived from the reflection of the same mode. The frequency axes were determined using the position of the sidebands that are required to generate the PDH signal.}
	\label{fig:spectrum-zoomin}
\end{figure}

A more detailed transmission spectrum around a principal resonance is shown in Fig.~\ref{fig:spectrum-zoomin}. The corresponding error signal which we obtained can be seen in the same figure.

Due to the finite rise time of our relatively slow photodiode (Thorlabs DET100A, $43~\mathrm{ns}$) it is difficult to get an accurate linewidth from this. First of all this means that the photodiode will keep showing a signal for a (short) while after the cavity has stopped doing so. Additionally, the cavity `bleeds' light for a while after the laser is scanned across resonance due to the buildup of light in it. We can make a crude estimate of the width of the cavity features, though: from Fig.~\ref{fig:spectrum-zoomin} we estimate the FWHM to be around one MHz; dividing $f_\mathrm{FSR}$ by the FWHM, we find that $\mathcal{F} \sim 1500$. This is sufficient for our present purposes, which is to lock the lasers and reduce their linewidth to $\ll 1~\mathrm{MHz}$.

\section{Linewidths \& stability}\label{sec:linewidths}
The final verdict of the performance of our cavity design is based on two indicators: the linewidth of the lasers that are locked to it, and the overall stability of the positions of the resonances. Both these values can be obtained using a heterodyning setup. The measurement of the long-term drift is discussed in Subsection~\ref{sec:cavity-drift}, but first we will focus on measuring the short-term laser linewidths. This can be achieved using a technique called `self heterodyning,' which involves splitting the light in to two arms, one of which is delayed for a time on the order of the laser coherence time $\tau_c$ or longer \cite{Okoshi80}. The resulting two beams can be considered as independent, and when they are overlapped on a photo diode they will produce a beating signal. The self-heterodyne technique measures the phase diffusion of the laser (on the timescale of the optical delay, see below), which is the dominant source of the laser linewidth for our lasers.

\subsection{Self-heterodyning}
To analyze the linewidths of our lasers, we used a self-heterodyning setup with a $12.4~\mathrm{km}$ long fiber. We used an AOM in double-pass configuration\footnote{By using it in a double-pass configuration it is easier to switch between a $780$ and $960~\mathrm{nm}$ beam since the AOM then does not need realigning.} to shift the other beam by $160~\mathrm{MHz}$. The beams are recombined on a fast photo diode (New Focus, 1601FS-AC), which is connected to a spectrum analyzer to measure the beat signal.

The fiber has a refractive index $n_f$ of approximately $1.5$ \cite{Rengelink13}, which means that the delay time $\tau_\mathrm{del.}$ is approximately $ 60~\mu\mathrm{s}$. Note that the refractive index depends on wavelength, so this is just an estimate. The delay time sets an upper bound for the linewidth that can be measured accurately, since $\delta\! f \propto 1/\tau_c$. If $\tau_c \gtrsim \tau_\mathrm{del.}$ the beams will still be (partially) coherent, meaning some interference will result.

If the two beams are fully incoherent the beating spectrum can be thought of as a convolution of the laser with itself. This means that the laser linewidth will be between a factor $1/\sqrt{2}$ and $1/2$ of the measured linewidth \cite{Okoshi80}, depending on whether we assume a Lorentzian or Gaussian lineshape.

A Voigt function is the most complete description \cite{Siegman86}; this is a convolution of a Gaussian and Lorentzian profile, and will result in a combined linewidth that is within the bounds set by its constituent distributions. It is suitable for situations in which two or more broadening sources with Lorentzian and Gaussian characters (e.g.\ the natural laser linewidth and technical phase noise) play a significant role.

Under the assumption of a white (frequency) noise distribution it is still possible to obtain a linewidth from a measurement in which some coherence remains. These measurements, however, are notoriously difficult to interpret accurately \cite{Bennetts14, vanExter92}. In those cases it is possible to fit an expression to the power spectrum of the form \cite{Horak06,vanExter92,Bennetts14}
\begin{align}\label{eq:powerfun}
	S(f) \propto& \frac{\delta\!f}{ \left( f - f_0 \right)^2 + \left( \delta\!f\right )^2 } \times	\notag \\
	&\Bigg( 1 - e^{-2\pi\delta\! f\tau_\mathrm{del}} \bigg\{ \cos\left[ 2\pi \tau_\mathrm{del}\left( f - f_0 \right) \right] + \notag \\
	&\qquad\qquad + \frac{\delta\! f}{f - f_0}\sin\left[ 2\pi \tau_\mathrm{del} \left(f - f_0\right) \right]\bigg\}\Bigg).
\end{align}

In the fully incoherent limit ($\delta\! f \tau \gg 1$) the exponent will go to zero, resulting in a purely Lorentzian lineshape. This expression also shows that, in the case of some remaining coherence, the fringes will have a period of $\tau_\mathrm{del}$, which we can compare to what we expect based on the length and refractive index of the fiber. Even though analyzing these fringes can be tricky, at the very least their presence provides a characteristic scale for the linewidth. The key is the suppression factor $\exp\left( -2\pi \delta\! f \tau_\mathrm{del} \right)$ which is non-negligible for $\delta\! f \lesssim 1/2\pi\tau_\mathrm{del}$; for us this last factor is approximately $3~\mathrm{kHz}$.

\subsubsection{960~nm laser}
In a self-heterodyning measurement we measure both the free-running laser spectrum, and the stabilized laser linewidth. For the latter the laser is locked to the cavity using a high-bandwidth PID controller (Toptica, FALC 110) which has a bandwidth of about $45~\mathrm{MHz}$ \cite{Toptica16}. This is an absolute prerequisite if one is to narrow a diode laser, which typically has frequency noise above the MHz level \cite{Zhao10}. We used the linewidth measurement to optimize the settings of the PID controllers by looking at how the profiles narrowed and developed fringes as we changed the settings of the several integrators that are in the controllers. What we present here are the results obtained using optimized settings.

\begin{figure}
	\centering
	\includegraphics[width=0.45\linewidth]{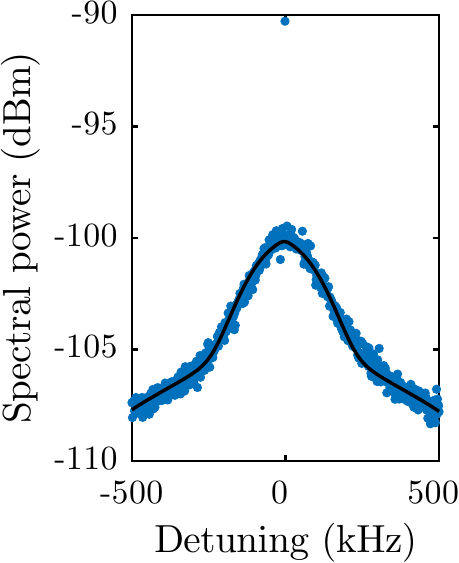}
	\includegraphics[width=0.45\linewidth]{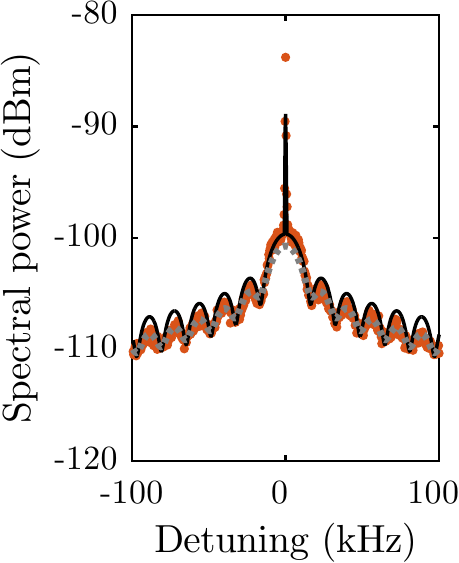}
	\caption{Self-heterodyned interferometric spectra of the unlocked (left) and locked (right) $960~\mathrm{nm}$ laser (measured with resolution bandwidths of $1~\mathrm{kHz}$ and $100~\mathrm{Hz}$, respectively). To fit the unlocked spectrum we used a pseudo-Voigt lineshape, yielding a laser linewidth of around $200~\mathrm{kHz}$. The fits in the locked spectrum are to Eq.~(\ref{eq:powerfun}), and are optimized for two different regimes: they fit to either the central region for  absolute detunings $\lesssim 10~\mathrm{kHz}$ (black, solid) or to the fringes, absolute detunings $\gtrsim 50~\mathrm{kHz}$ (grey, dashed), and lead to laser linewidths $\delta\! f$ of $4$ and $20~\mathrm{kHz}$, respectively.}
	\label{fig:960-beat-sigs-fitted}
\end{figure}

The beat signal was analyzed using a Rigol DSA815 spectrum analyzer. Each measurement consists of about 30 averages to increase the signal-to-noise ratio. Figure~\ref{fig:960-beat-sigs-fitted} contains both the locked and unlocked spectra of the $960~\mathrm{nm}$ laser. To obtain quantitative information from these results we fitted several functions to them, the fits are included in Fig.~\ref{fig:960-beat-sigs-fitted} as well. For the unlocked laser spectrum we used a pseudo-Voigt profile, which is a weighted sum of a Lorenzian and a Gaussian function \cite{Wertheim74,Liu01}. From this we obtained a Lorentzian (FWHM) linewidth of $70~\mathrm{kHz}$ and a Gaussian (FWHM) linewidth of $160~\mathrm{kHz}$. The fitted pseudo-Voigt was 18\% Lorentzian, and 82\% Gaussian. The $-3~\mathrm{dB}$ points are $320~\mathrm{kHz}$ apart, which means that the laser's linewidth is between $160$ and $230~\mathrm{kHz}$, depending on its spectral character (see above, also Ref.~\cite{Okoshi80}). There exists an empirical expression to obtain the FWHM of a Voigt profile ($\gamma_V$) from its constituent Lorentzian and Gaussian profiles' widths ($\gamma_L$ and $\gamma_G$, respectively) \cite{Olivero77}. It is given by
\begin{equation}
	\gamma_V = 0.5346 \gamma_L + \sqrt{0.2166 \gamma_L^{~2} + \gamma_G^{~2}}.
\end{equation}
With our fit parameters this results in a Voigt linewidth of $200~\mathrm{kHz}$ which is in accordance with what we expect based on the distance between the $-3~\mathrm{dB}$ points.

The locked laser spectra were fitted using Eq.~(\ref{eq:powerfun}). As is pointed out in Ref.~\cite{vanExter92}, it is difficult to get this function to fit both the fringes and the central part; we follow an approach similar to that taken in said reference where they fit two functions (each with the same total spectral intensity): one that matches the modulation depth of the fringes, and one that  matches the power level of the wings. The reason that it is not possible to make a single good fit is that that the phase noise is not white, as is assumed in Eq.~(\ref{eq:powerfun}), but has some additional low-frequency components which are not accounted for \cite{vanExter92}. For the 960~nm laser we chose to fit the central peak and fringes separately.

\begin{figure}
	\centering
	\includegraphics[width=0.45\linewidth]{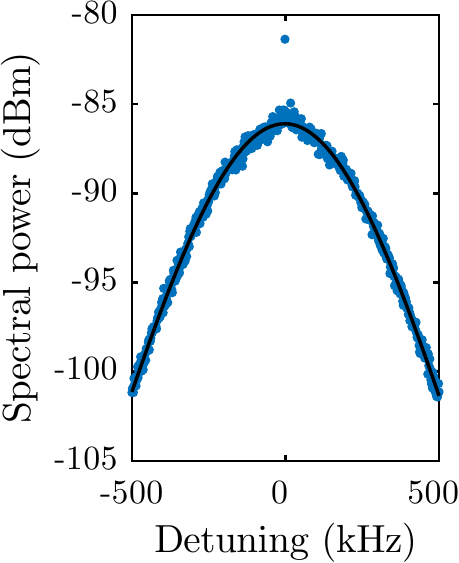}
	\includegraphics[width=0.45\linewidth]{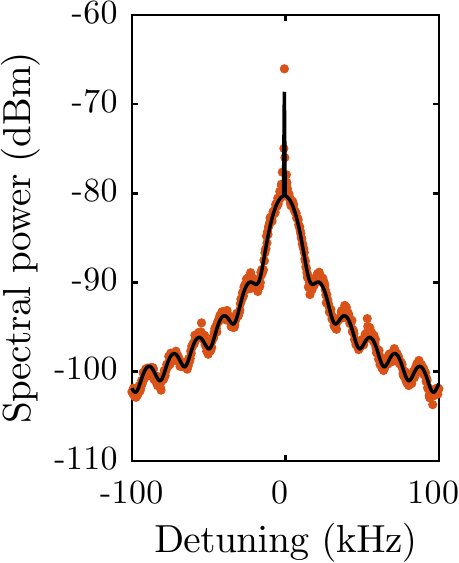}
	\caption{Self-heterodyned interferometric spectra of the unlocked (left) and locked (right) $780~\mathrm{nm}$ laser (measured with resolution bandwidths of $1~\mathrm{kHz}$ and $100~\mathrm{Hz}$, respectively). We used a Gaussian lineshape to fit the unlocked spectrum; this led to a laser linewidth of $300~\mathrm{kHz}$. To fit the locked spectrum we used Eq.~(\ref{eq:powerfun}) again, this yielded a linewidth of $\delta\! f$ of $3~\mathrm{kHz}$.}
	\label{fig:780-beat-sigs-fitted}
\end{figure}

The delta-peak in all these measurements is caused by the AOM being aligned imperfectly. An unshifted portion of light is still present in the path that is not delayed, leading to a sharp peak at the AOM frequency regardless of whether the delayed light is present or not. The peak is very narrow, but in practice it is broadened by the finite resolution bandwidth of the spectrum analyzer. We account for it by adding a delta peak at zero frequency to the fitting function, but this is only done to prevent our fitting routine from attempting to fit Eq.~(\ref{eq:powerfun}) to the narrow feature.

We obtain linewidths $\delta\! f$ of $4$ and $20~\mathrm{kHz}$ when fitting to the central part and the wings, respectively. Both fits result in a similar value for the fringe spacing, corresponding to a delay time of $61.9~\mu\mathrm{s}$, which is in accordance with what we expect based on the fiber specifications.

\subsubsection{780~nm laser}
To measure the linewidth of the $780~\mathrm{nm}$ laser we largely followed the same procedure as for the $960~\mathrm{nm}$ laser. The difference between the locked and unlocked laser is shown in Fig.~\ref{fig:780-beat-sigs-fitted}. These data were fitted as well (see Fig.~\ref{fig:780-beat-sigs-fitted}); for the unlocked laser a simple Gaussian profile fitted sufficiently well, and we obtained a FWHM laser linewidth of $300~\mathrm{kHz}$. For the locked laser we found a $\delta\! f$ of $3~\mathrm{kHz}$, and a delay time of $62.2~\mu\mathrm{s}$ (a single function fit the profile sufficiently well). The difference between the delay times between the two lasers (as far as it is outside of the measurement error) can be explained by a wavelength-dependence in the index of refraction of the fiber. This last result for $\delta\! f$ should certainly be taken with a grain of salt for reasons mentioned above (i.e.\ the assumption of the noise character), although it should be pointed out that for this measurement it is possible to fit Eq.~(\ref{eq:powerfun}) to both the central region and the fringes. (This was difficult for the $960~\mathrm{nm}$ laser, cf.\ Fig.~\ref{fig:960-beat-sigs-fitted}.)

\subsection{The cavity drift}\label{sec:cavity-drift}
In our experiments involving Rydberg excitation we have noticed an apparent drift in the resonance frequency over time. This is caused by a drift in the frequency of the cavity feature to which the lasers are referenced.

To measure the cavity drift we used our $780~\mathrm{nm}$ Rydberg laser, which was heterodyned with a beam derived from another $780~\mathrm{nm}$ laser in our laboratory, which is locked to a rubidium vapor cell via Doppler-free saturated absorption spectroscopy. We used the same heterodyning setup as we used for the linewidth measurements described above, but without the delay fiber. This means that one of the two lasers is shifted by an additional $160~\mathrm{MHz}$ by the AOM that is still on the board.

The reference laser is locked to a rubidium vapor cell, so it is relatively broad but stable over time. We measured its linewidth to be $360~\mathrm{kHz}$ (full-width at half maximum).

This linewidth is good enough to monitor frequency drifts on the order of MHz. The reference laser is locked to the $F = 2 \rightarrow F'=1,3$ crossover, and the Rydberg $780~\mathrm{nm}$ laser is blue-detuned\footnote{This seemingly arbitrary number was set by the cold-atom Rydberg experiments we were running in parallel.} by $300~\mathrm{MHz}$ from the $F = 2 \rightarrow F' = 3$ transition. Therefore the frequency difference between the lasers is $(211.8 + 300)~\mathrm{MHz} = 511.8~\mathrm{MHz}$ \cite{Steck08}. The AOM still in the setup shifts the Rydberg laser by an additional $160~\mathrm{MHz}$. This means that the frequency difference between the two lasers is approximately $670~\mathrm{MHz}$.

We measured the drift by reading out the peak beat frequency from the spectrum analyzer every ten seconds. Using this and our relatively stable laser locks we were able to do measurements over multiple days.

\begin{figure}
	\centering
	\includegraphics[width=0.45\textwidth]{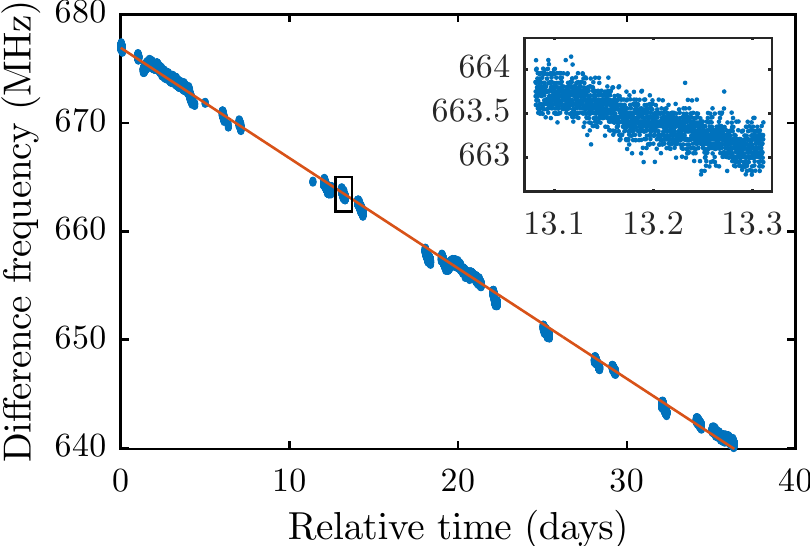}
	\caption{Characterization of the long-term drift of the cavity through measurements of the difference frequency between two lasers over the course of a month. The linear fit has a slope of $-1.02~\mathrm{MHz/day}$. Inset: data from a single day (covering $\approx 5$ hours) indicated by the rectangle.}
	\label{fig:cavity-drift-month}
\end{figure}

Figure \ref{fig:cavity-drift-month} shows a combination of such measurements taken over the course of a month.

When the system reached equilibrium after having switched on the stabilization circuit for the mounts's temperature as well, an overall drift was left; see Fig.~\ref{fig:cavity-drift-month}. This was about $1~\mathrm{MHz/day}$.

We can estimate how much the cavity has to expand by to achieve a frequency shift of $1~\mathrm{MHz}$. For a wavelength of $780~\mathrm{nm}$ this is $0.26~\mathrm{nm}$.

This apparent expansion may be related to our mirrors being glued rather than optically contacted; this is an important difference between this design and those presented elsewhere \cite{SLSdata,Alnis08,Ludlow07,Zhang13,Wu16}. A possible mechanism for this slow expansion could be that the glue crept between the spacer and the mirrors, and is still curing, resulting in the length changes we appear to be seeing.

The effect of temperature changes on the spacer was measured by disabling the circuit that regulates the temperature of the cavity and its mounts. This led to an instantaneous increase of coupling in of laboratory temperature fluctuations. Specifically, we noticed that an sudden increase of $1~\mathrm{K}$ in laboratory temperature (i.e.\ as happens after turning on all laboratory equipment) led to a gradual drift of $\sim5~\mathrm{MHz}$ in resonance frequency over a period of 10 hours. Using these data in conjunction with Eq.~(\ref{eq:freq-shift-expansion-coeff}) and the estimated thermal RC time of 17 hours leads to an expansion coefficient $\alpha$ of $23\cdot 10^{-9}~\mathrm{K}^{-1}$, which is inside the specified range \cite{ULEData}.

\section{Discussion}
The cavity and the laser system locked to it are now being successfully applied to excite Rydberg atoms and perform Rydberg spectroscopy of ultracold rubidium atoms in our laboratory \cite{Cisternas17}.

For a next-generation design it would be commendable to assess the thermal coupling between the vacuum system surrounding the cavity and the surface on which it is mounted in greater detail. In our case the interface is that of stainless steel against stainless steel, which has an excellent thermal conductivity. One of the advantages is that this does make the system mechanically stable, and that the alignment of the laser beams into the cavity only rarely needs to be adjusted. The timescale of the observed thermal fluctuations is also important, because if they are slow and persistent enough they will couple into the system at some point anyway, regardless of any shielding between mounts and mounting surface. This means active thermal feedback is essential; it might be very beneficial to add this to the inside of the vacuum system, although this would involve designing a vacuum-compatible regulating circuit. By adding thermal regulation to the vacuum system's exterior as well one should be able to achieve a very stable system. In that case the only drift that will remain is the one that seems to be inherent to the assembly of spacer and mirrors. For this to be minimized, it would also be highly recommended to optically contact the mirrors to the spacer.

If one is to use this design as a template for a cavity of a higher finesse, this can be done (up to a point) by increasing the reflectivity of the mirrors. One should also bear in mind that the performance of high-finesse cavities is in part limited by their sensitivity to vibrations; our design was not optimized in this regard, and it may be necessary to perform such an optimization if higher stability is desired.

\section{Conclusion}\label{sec:conclusion}
We have demonstrated an economical design for a medium-finesse, dichroic, optical cavity ($\mathcal{F}\sim 1500$). Using a heterodyning setup, we measured laser linewidths of  $\delta\! f\lesssim 10~\mathrm{kHz}$ for both of our Rydberg lasers, and a long term drift of about $1~\mathrm{MHz/day}$ for our $780~\mathrm{nm}$ laser. The cavity was protected from thermal fluctuations of the environment using two independent temperature regulation circuits. The resulting system is now in active use to stabilize lasers for the excitation of ultracold rubidium atoms to Rydberg states.

\section*{Acknowledgments}
We thank Alex Bayerle and Florian Schreck for stimulating discussions concerning our design, and H.B. van Linden van den Heuvell for a critical reading of the manuscript.

This work is financially supported by the Foundation for Fundamental Research on Matter (FOM), which is part of the Netherlands Organisation for Scientifc Research (NWO). We also acknowledge financial support by the EU H2020 FET Proactive project RySQ (640378).

\bibliography{bib}

\end{document}